\documentclass[12pt]{article}
\usepackage[dvips]{epsfig}
\usepackage{overcite,dcolumn,amsmath}
\newcommand{\figwidth}{\linewidth}

\newcommand{\figcaption}[1]{\caption{#1}}
\newcommand{\caplist}[1]{}

\begin{document}
\title{Local Reorientation Dynamics of Semiflexible Polymers in the Melt}
\author{Roland Faller$^1$ \and Florian M\"uller-Plathe$^{1,*}$ 
  \and Andreas Heuer$^{1,2}$\\
  \small $^1$ Max-Planck-Institut f\"ur Polymerforschung, D-55128 Mainz, 
  Germany\\
  \small $^2$ Universit\"at M\"unster, Institut f\"ur Physikalische Chemie,\\
  \small Schlossplatz 4/7, D-48149 M\"unster, Germany}
\maketitle

{\it For submission to Macromolecules}\\

\abstract{
  \noindent The reorientation dynamics of local tangent vectors of chains in
  isotropic amorphous melts containing semiflexible model polymers was studied
  by molecular dynamics simulations. The reorientation is strongly influenced
  both by the local chain stiffness and by the overall chain length. It takes
  place by two different subsequent processes: A short-time non-exponential
  decay and a long-time exponential reorientation arising from the relaxation
  of medium-size chain segments. Both processes depend on stiffness and chain
  length. The strong influence of the chain length on the chain dynamics is in
  marked contrast to its negligible effect on the static structure of the
  melt. The local structure shows only a small dependence on the stiffness,
  and is independent of chain length. Calculated correlation functions related
  to double-quantum NMR experiments are in qualitative agreement with
  experiments on entangled melts. A plateau is observed in the dependence of
  segment reorientation on the mean-squared displacement of the corresponding
  chain segments. This plateau confirms, on one hand, the existence of
  reptation dynamics. On the other hand, it shows how the reptation picture
  has to be adapted if, instead of fully flexible chains, semirigid chains are
  considered.
}
\section{Introduction}
\label{sec:intro}
Modern double-quantum nuclear magnetic resonance (NMR) experiments aim to
understand the microscopic dynamics of polymer segments in the
melt.\cite{schmidt-rohr94} The polymer dynamics is most often described by
either the Rouse\cite{rouse53} or the reptation\cite{degennes71,doi86} model
depending on chain length. Both models, however, take the architecture of a
specific polymer into account only via the so-called Kuhn length. These local
features are summarized into
$C_{\infty}=\frac{\langle\vec{R}^2\rangle}{Nl_b^2}$, the ratio between the
mean squared end-to-end distance $\langle\vec{R}^2\rangle$ and the number of
monomers $N$ (multiplied with the squared bond length $l_b^2$). Statics and
dynamics are thus renormalized onto a Gaussian bead-spring chain of larger but
fewer beads of the size of the Kuhn length
$l_K=\frac{\langle\vec{R}^2\rangle}{Nl_b}$. Experimentally, however, there is
a qualitative difference between fully flexible polymers like
poly-(dimethyl-siloxane) (PDMS)\cite{callaghan98} and moderately stiff systems
like poly-(butadiene) (PB)\cite{graf98} when it comes to the reorientation of
chain segments. These differences remain after both polymers have been
appropriately renormalized onto Gaussian chains. The PB system shows unusual
dynamics which has been taken as an indication of local order\cite{graf98},
whereas the PDMS data is quite successfully described by the standard
reptation picture\cite{callaghan98}. From the shape of reorientation
auto-correlation functions a relatively high degree of residual structural
order in the presence of entanglements has been deduced for PB. The
experimental data, however, cannot be interpreted without model
assumptions. Here, simulations which are validated against experimental raw
data may contribute to a better understanding.\cite{mplathe00s}

Molecular dynamics simulations are widely applied to study the dynamics of
simplified polymer models in solution\cite{duenweg91,duenweg93} and the
melt\cite{kremer88,rigby88,kremer90,takeuchi91,duenweg98,faller99d,puetz00}.
To date, only simple models are capable of investigating dynamics of long
entangled polymer chains in the melt to a satisfactory degree of
accuracy. Models that allow for atomistic details are limited to shorter times
and to fewer and shorter chains.\cite{moe95,paul95,harmandaris98,mplathe98}

Recently, we showed that the {\it static} local mutual orientation of
neighboring chains depends on chain stiffness in the amorphous melt, whereas
the chain length has no influence on local static
properties.\cite{faller99d,faller99b} In the present contribution, we now
extend our investigations to the {\it dynamics} of entangled and unentangled
melts of polymer chains with local stiffness. In the following section, the
polymer model is shortly recapitulated and details of the simulated systems
are described. In the main part (Section~\ref{sec:reorient}), reorientation
correlation functions are analyzed and compared to theoretical considerations
and experiments. Section~\ref{sec:structure} relates static chain packing and
dynamic chain reorientation observables.
\section{Model and Computational Details}
We performed polymer melt simulations using Brownian dynamics (BD) of a widely
used and well characterized generic polymer
model\cite{kremer88,kremer90,duenweg98} with added local
stiffness.\cite{faller99b} All monomers interact via a truncated and
shifted, therefore purely repulsive, Lennard-Jones potential
(Weeks-Chandler-Andersen, WCA, potential\cite{weeks71})
\begin{equation}
  V_{WCA}(r)=4\epsilon\left[\left(\frac{\sigma}{r}\right)^{12}-
    \left(\frac{\sigma}{r}\right)^{6}+\frac{1}{4}\right],\;
  r<\sqrt[6]{2}\sigma\;.
\end{equation}
Neighbors on the chain are connected by a finitely extensible
non-linear elastic (FENE) potential which is used for computational efficiency
\begin{eqnarray}
  V_{FENE}(r)&=&-\frac{\alpha}{2}\frac{R^{2}}{\sigma^{2}}
  \ln\left(1-\frac{r^{2}}{R^{2}} \right),\\
  r<R&=&1.5\sigma\;,\quad\alpha =30\epsilon\;.\nonumber
\end{eqnarray}
This yields, together with the WCA-potential, an anharmonic spring.  Most of
our systems have an additional three-body potential to stiffen the chain
locally 
\begin{equation}
  V_{bend}=b\left(1-\frac{\vec{r}_{i-1,i}\cdot{\vec{r}_{i,i+1}}}
    {r_{i-1,i} r_{i,i+1}}\right)\;,
  \quad \vec{r}_{ij} := \vec{r}_{j}-\vec{r}_{i}\;.
  \label{eq:bend}
\end{equation}
For details of the implementation and parallelization, see
ref.~\citen{puetz98}. Throughout this work, reduced units are used with
mass~$m$, bead diameter~$\sigma$, and the strength of the WCA
potential~$\epsilon$ set to unity. The time unit
is~$t^*=\sigma\sqrt{\frac{m}{\epsilon}}$. Temperature is measured in units
of~$\epsilon$ by setting Boltzmann's constant~$k_{B}=1$. The average bond
length is~$l_{b}=0.97\sigma$ so that the beads overlap only slightly.

Systems of up to 1000 chains of length $N$ monomers ranging from 5 to 1000
were simulated at melt density ($\rho^{*}=0.85\sigma^{-3}$) and temperature
$T^{*}=1$. The Brownian dynamics algorithm is mainly used to maintain this
temperature. Additionally, it has been shown to be very efficient for our
model~\cite{kremer90}. The monomer friction coefficient of 0.5 inverse time
units used here means that only processes on a time scale well above one of our
time units have a meaningful dynamics. The persistence lengths $l_{p}$ defined
via the decay of the bond direction correlation function along the chain
backbones varied from one bond length up to 5 bond lengths.
\begin{equation}
  \Big\langle\vec{u}(0)\vec{u}(l)\Big\rangle=e^{-l/l_{p}},
\end{equation}
The distance $l$ is measured along the contour of the chain. The persistence
length is related to the Kuhn length (Section~\ref{sec:intro}) in the
worm-like chain model by $l_p=\frac{1}{2}l_K$.\cite{doi86} In the simulations,
$l_p$ is controlled by choosing appropriate values of $b$ in
Eq.~\ref{eq:bend}. Hence, throughout this article we refer to systems of
different stiffness with their $l_p$ rather than the corresponding $b$. The
most flexible system ($l_{p}\approx l_{b}$) has no intrinsic stiffness, only
the excluded-volume interaction leads to its persistence length being
non-zero. The persistence length increases weakly with chain length due to end
effects.\cite{faller99b} The values indicated here are the limits for long
chains. For $N\ge 10$ the persistence length is much shorter than the contour
length. An overall Gaussian behavior is therefore expected and confirmed by
the characteristic ratio
$\lim_{N\to\infty}\frac{R^{2}_{end-end}}{R^{2}_{gyr}}\approx
6$.\cite{faller99b} Note that the model would show a nematic liquid
crystalline phase if the bending stiffness was increased far above
$l_p=5$.\cite{kolb99}

All chains of length up to 75 and all flexible chains $(l_p=1)$ relaxed fully,
as evidenced by the decay of all the Rouse modes. To cut down necessary
equilibration times, the longer chains (or chains of greater stiffness) were
initialized as non-reversal random walks whose local structure was estimated
from simulations of shorter chains: In the setup configuration a monomer $i$
and its second neighbor $i+2$ are not allowed to approach closer than a
certain distance. This setup procedure reduces the equilibration time
substantially while producing useful configurations.\cite{kremer90}

The end-to-end distance and the gyration radius changed then only very
slightly in the initial stage of the simulation. Their equilibrium values as
a function of stiffness were already presented in reference~\citen{faller99b}
together with other static observables like structure functions, pair
distribution functions and local chain orientation correlation functions. Some
of this data is included in section~\ref{sec:structure}.  It is not yet
possible to simulate the ``slowest'' systems (e.g. $l_{p}=5,\quad N=1000$)
until the final regime of free diffusion is reached. Still, we trust that even
this system is sufficiently equilibrated for the purpose under study.

\begin{figure}
  \includegraphics[angle=-90,width=\figwidth]{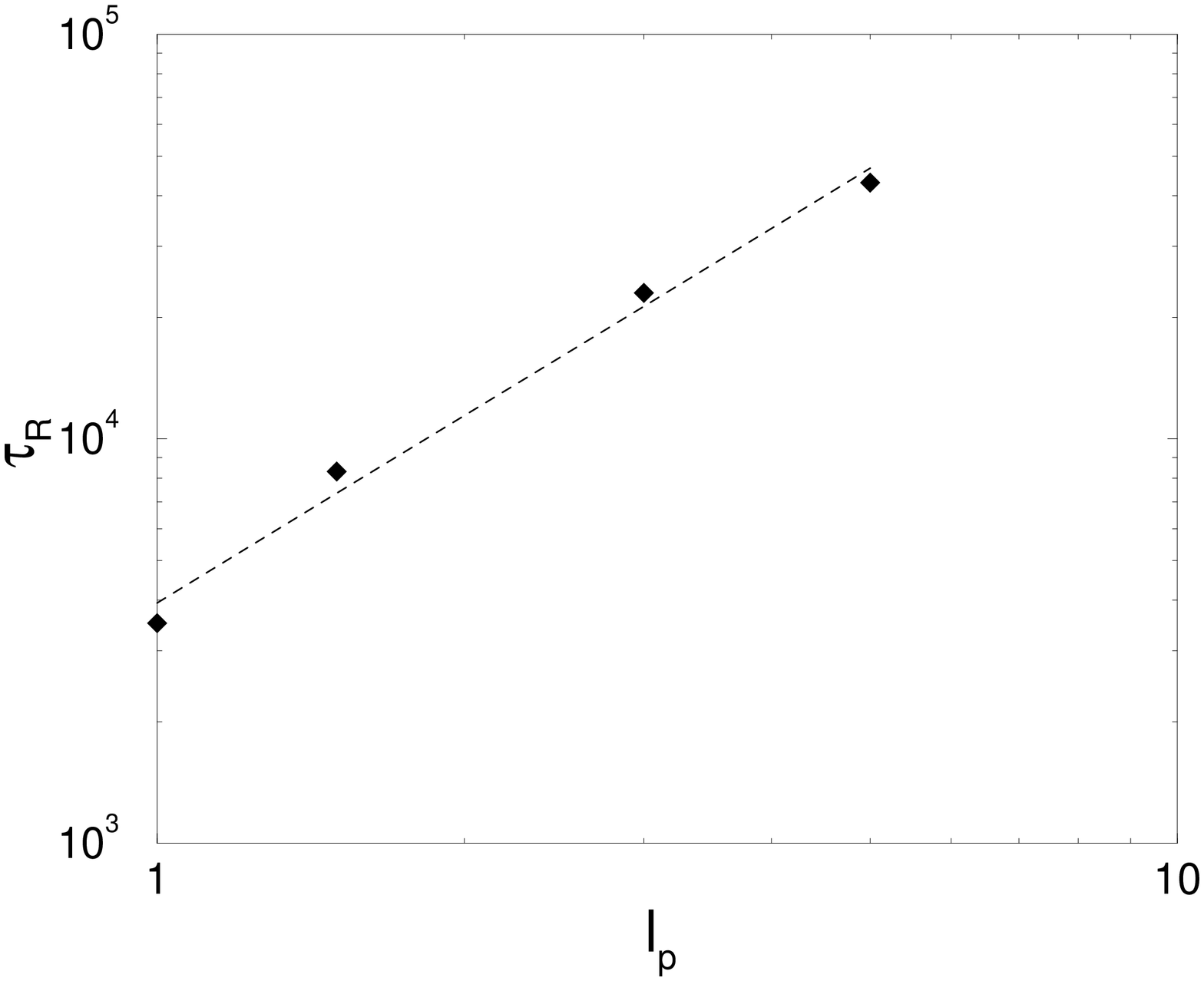}
  \figcaption{Increase of Rouse times with persistence length at $N=50$. The
    dashed line corresponds to an algebraic increase with $l_p^{1.54}$.}
  \label{fig:norouse}
\end{figure}

Table~\ref{tab:systems} gives an overview of the simulated systems with their
respective Rouse times and simulated times. We are aware that for non-flexible
polymers the Rouse modes are no longer the true eigenmodes (see
section~\ref{sec:reorient}). Still, the Rouse time is useful as an estimate of
the 
relaxation time. For chains of equal length, the Rouse time $\tau_R$ would
increase linearly with chain extension, i.e. $l_p$, if the friction due to
neighboring chains were constant.\cite{doi86} However, the relaxation times
increase even stronger (Table~\ref{tab:systems} and
Figure~\ref{fig:norouse}). Similarly, for the non-flexible chains ($l_p>1$),
the increase of $\tau_R$ with $N^{2}$ as expected from the Rouse model is no
longer observed. The slowdown is stronger, indicating an earlier onset of
entanglement influence with increasing persistence length
(Table~\ref{tab:systems}).
\begin{table}
  \begin{center}
    \[
    \begin{array}{|D{.}{.}{-1}|r|r|r|r|D{.}{.}{-1}|}
      \hline
      l_{p} & N & N_{c} & t_{sim} & \tau_R & 6DN\\
      \hline
      1.4 & 5 & 1000 & 30000 &  (40) & 0.44 \\
      3.0 & 5 & 1000 & 30000 &  (62) & 0.42 \\
      5.0 & 5 & 1000 & 30000 &  (88) & 0.38 \\
      \hline
      1.0 & 10 & 500 & 30000 & 140 & 0.43 \\
      1.4 & 10 & 500 & 30000 & 170 & 0.43 \\
      3.0 & 10 & 500 & 30000 & 290 & 0.35 \\
      5.0 & 10 & 500 & 33000 & (730) & 0.27 \\
      \hline
      1.4 & 13 & 500 & 25000 & 290 &  0.35\\
      \hline
      1.4 & 15 & 500 & 25000 &  370 &  0.35\\
      \hline
      1.4 & 17 & 500 & 67000 &  590 & 0.30 \\
      3.0 & 17 & 500 & 30000 & 1070 & 0.27 \\
      5.0 & 17 & 500 & 30000 & 2600 & 0.21 \\
      \hline
      1.4 & 20 & 500 & 30000 &  780 & 0.29 \\
      3.0 & 20 & 500 & 20000 & 1700 & 0.23 \\
      5.0 & 20 & 500 & 25000 & 3700 & 0.19 \\
      \hline
      1.0 & 25 & 500 & 30000 &  900 & 0.42 \\
      1.4 & 25 & 500 & 25000 & 1400 & 0.29 \\
      3.0 & 25 & 500 & 63000 & 2900 & 0.20 \\
      5.0 & 25 & 500 & 78000 & 6400 & 0.18 \\
      \hline
      1.4 & 30 & 500 & 25000 & 2000 & 0.26\\
      3.0 & 30 & 500 & 25000 & 5000 & 0.17\\
      \hline
      5.0 & 35 & 500 & 53000 & 21000 &\\
      \hline
      1.0 & 50 & 500 & 20000 &  3500 & 0.31 \\
      1.4 & 50 & 500 & 22000 &  8300 & 0.18 \\
      3.0 & 50 & 500 & 70000 & 23000 & 0.10 \\
      5.0 & 50 & 500 & 50000 & 43000 & \\
      \hline
      1.4 & 75 & 500 & 76000 & 27000 & 0.13 \\
      \hline
      1.4 & 200 & 500 & 300000 & ^{*}300000 & \\
      3.0 & 200 & 500 & 350000 & ^{*}400000 & \\
      5.0 & 200 & 500 & 700000 & ^{*}700000 & \\
      \hline
      1.0 & 350 & 120 & 1700000 & 530000 & \\
      \hline
      5.0 & 1000 & 250 & 60000 & O(10^{7}) &\\
      \hline
    \end{array}
    \]
    \caption{Simulated systems: $l_{p}$: persistence length in monomers; $N$:
      chain length in monomers; $N_{c}$: number of chains, $t_{sim}$:
      Simulation times; $\tau_R$: Rouse times (obtained by exponential
      fit of the lowest 3 to 5 Rouse modes. $^{*}$ partly
      estimated), numbers in brackets for systems where chains are too short;
      $D$: center of mass diffusion constant (for the systems 
      where free diffusion could be reached). Some data of the system with 120
      chains of length 350 are also presented in ref.~\citen{puetz00}. The
      errors in $\tau_R$ are at most 5\% and the errors in $6DN$ are about
      0.01.} 
    \label{tab:systems} 
  \end{center}
\end{table}

The very short chains ($N\le15$) allow an estimate of the diffusion
coefficient of chains of a given stiffness, i.e. their mobility in the absence
of entanglements. The diffusion coefficient does not decrease linearly with
chain stiffness as would be expected from the Rouse model; for reptation this
decrease is quadratic. As entanglement length we take the chain length for
which the crossover from linear to quadratic takes
place.\cite{faller00sa,faller00a}
\section{Chain Reorientation}
\label{sec:reorient}
\subsection{Reorientation Correlation Function}
The main purpose of this work is to investigate the reorientation dynamics of
local chain segments in dense melts. This was studied by means of the
auto-correlation function of the second Legendre polynomial $P_2$ of chain
tangent vectors
\begin{equation}
  C_{reor}(t)=\Bigg\langle\frac{1}{2}\Big[3(\vec{u}(t)\vec{u}(0))^2-1\Big]
  \Bigg\rangle
  = \Big\langle P_2[\vec{u}(t)\vec{u}(0)]\Big\rangle
  \label{eq:creor}
\end{equation}

As chain tangent vectors we take (normalized) vectors connecting
neighboring beads
\begin{equation}
  \vec{u}=\frac{\vec{r}_{i+1}-\vec{r}_{i}}{|\vec{r}_{i+1}-\vec{r}_{i}|}
  \label{eq:unit}
\end{equation}
unless noted otherwise. The second, rather than the first, Legendre polynomial
is taken because its Fourier transform relates directly to NMR 
measurements ($T_{1}$ experiments) Also the double-quantum experiments aimed
at the chain dynamics are related to this function (below).
\subsubsection{Short-time behavior}
For all systems investigated, we have found that the reorientation correlation
function (Eq.~\ref{eq:creor}) consists of two qualitatively different
parts. For short times, its decay follows a power law (algebraic). At long
times, the decay is exponential. All characteristics of the reorientation
correlation function are influenced by the chain length $N$ and the chain
stiffness $l_p$: the short-time part, the long-time part, the time at which the
cross-over from algebraic to exponential behavior occurs and the function
value at this point. We will see that the short-time regime is influenced more
by stiffness, while the long-time regime shows a stronger dependence on chain
length. 

\begin{figure}
  \includegraphics[angle=-90,width=0.5\figwidth]{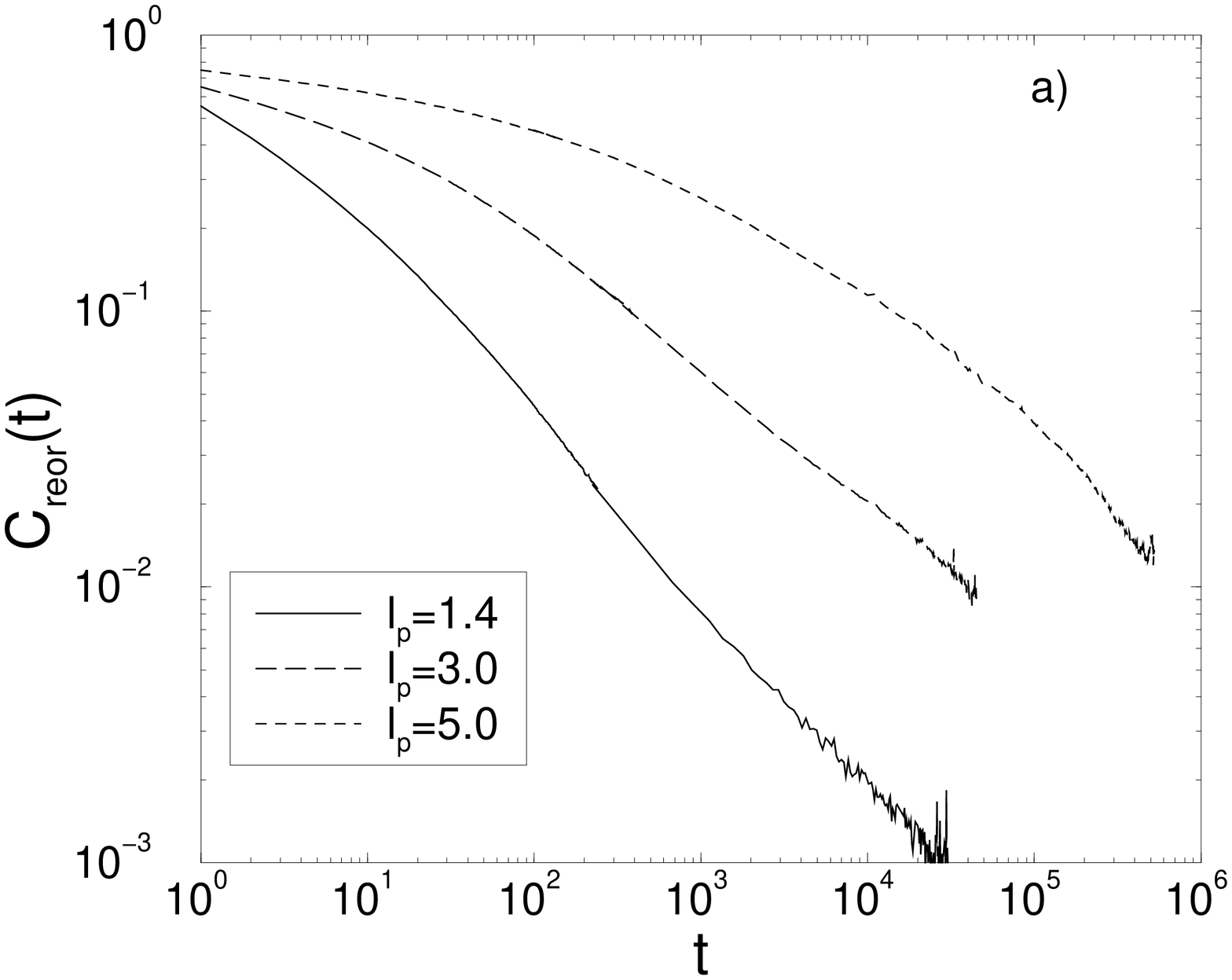}
  \includegraphics[angle=-90,width=0.5\figwidth]{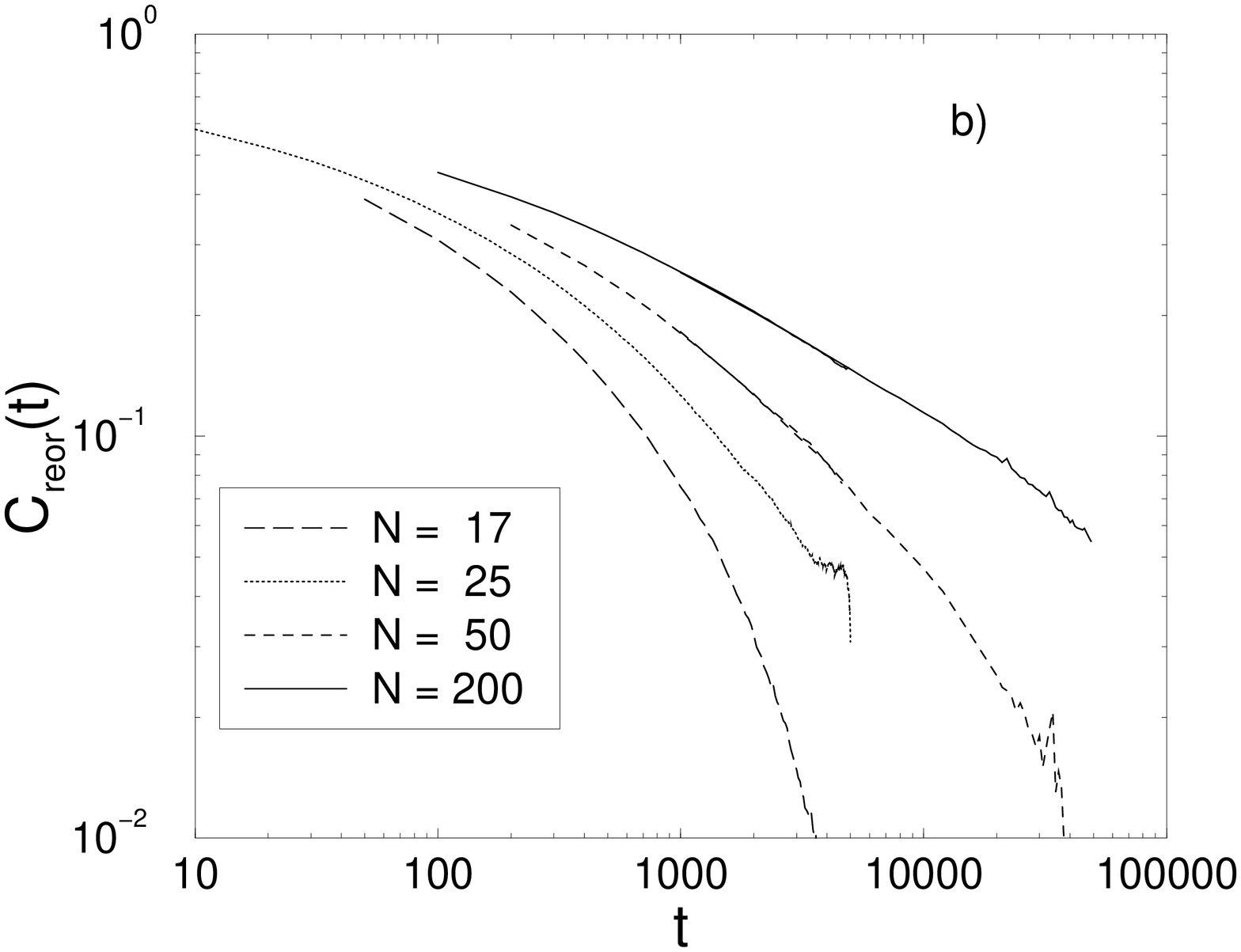}
  \includegraphics[angle=-90,width=0.5\figwidth]{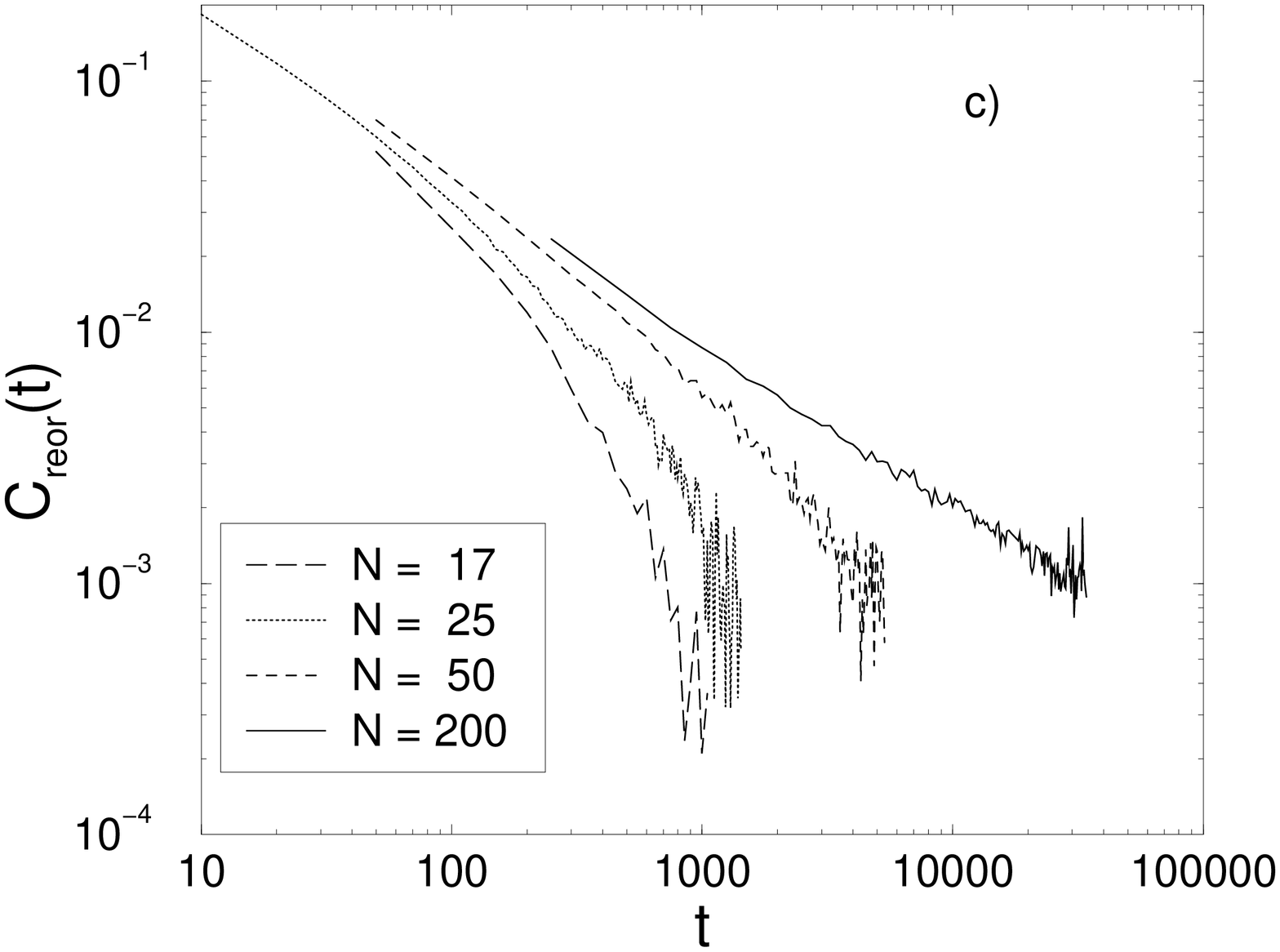}
  \includegraphics[angle=-90,width=0.5\figwidth]{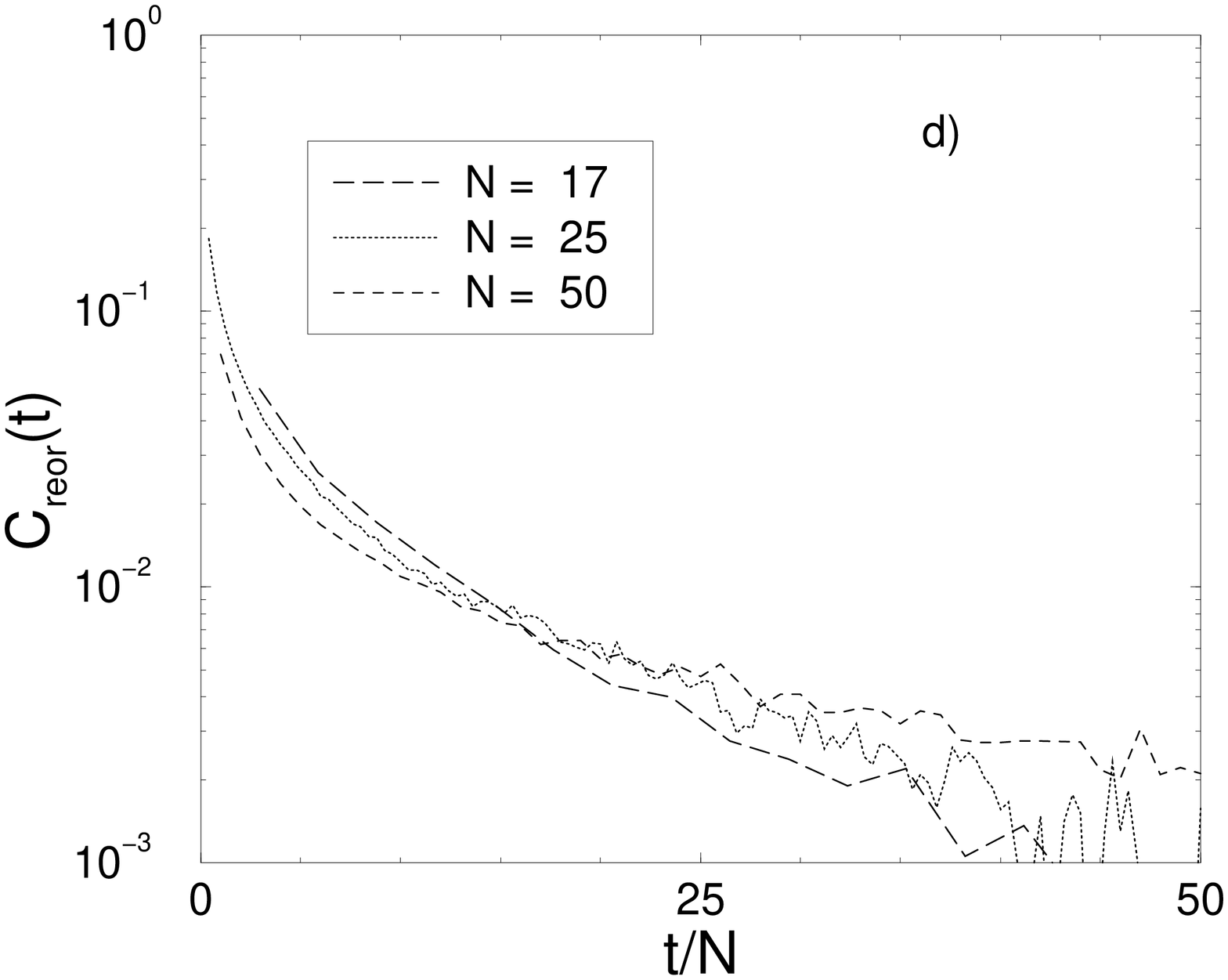}
  \figcaption{Short-time behavior of time dependent second Legendre polynomial
    of next neighbor vectors for different systems. a) $N=200$ different
    persistence lengths, b) $l_{p}=5$ different chain lengths, c)
    $l_{p}=1.4$ different chain lengths. d) System $l_{p}=1.4$ for very short
    times rescaled by $\frac{1}{N}$.}
  \label{fig:decay-algebra}
\end{figure}

The influence of the chain architecture on the short-time behavior is
illustrated in Fig.~\ref{fig:decay-algebra}. Reorientation is slowed by
increasing the stiffness at constant length (Fig.~\ref{fig:decay-algebra}a) as
well as by increasing the length at constant stiffness
(Figs.~\ref{fig:decay-algebra}b and ~\ref{fig:decay-algebra}c for $l_p=5$ and
$l_p=1.4$, respectively). The more flexible chains and the shorter chains
reach exponential behavior earlier. At the same time, the short-time algebraic
process is more "efficient" for the shorter and the more flexible chains,
i.e. the reorientation correlation function has decreased to a smaller value
when long-time exponential behavior sets in.
These findings can be understood if one attributes the short-time behavior to
local dynamics and the long-time behavior to the relaxation of larger chain
segments or, ultimately, to the rotational diffusion of entire chains. If the
chains are flexible much of the reorientation can be achieved by local
rearrangement, i.e. without the local tangent vector feeling that it is part
of a long entangled chain. As stiffness increases, it hinders the
reorientation on local scales, so the reorientation of the local tangent
vector has to wait for some larger-scale reorientation which is exponential.
The early crossover to exponential behavior and the apparently small
efficiency of the short-time process for short chains, on the other hand, is
due to a faster rotational diffusion of the entire chains, as they become
shorter. The rotational diffusion begins to contribute substantially to the
reorientation, before the local process can complete. Its time dependence is
exponential according to e.g. the Debye model.\cite{mcquarrie76}

If the Rouse model were strictly applicable the relaxation time of entire
chains in the melt (Rouse time $\tau_R$) should scale with $N^2$. Hence, for
constant stiffness the reorientation correlation functions belonging to
different $N$ should coincide if the time axis is transformed $t\to t/N^2$.
Instead, we find empirically that coincidence is achieved for $t\to t/N$
(Fig.~\ref{fig:decay-algebra}d). This indicates that even for a small
deviation from full flexibility ($l_p$=1.4) the Rouse model is
not appropriate for the short-time relaxation, the local process being
dominated by effects other than connectivity, e.g. stiffness. Additionally the
``bead friction'' enters which also increases sublinearly with
stiffness.\cite{faller00a} 

In contrast to the local dynamics, there is no chain length influence at all
on local structural properties.\cite{faller99b}

As a general result, we note that, while the short-time process is influenced
by both stiffness and chain length, the influence of stiffness is much
stronger. (This can, for example, be seen by comparing
Figures~\ref{fig:decay-algebra}b and~\ref{fig:decay-algebra}c.) It is,
therefore, to be expected that also in experiments on real polymers the
short-time regime will experience the influence from the chemical architecture
of the polymer, whereas the chain length should be secondary.

\subsubsection{Long-time behavior}
\label{sec:longtime}
\begin{figure}
  \includegraphics[angle=-90,width=0.5\figwidth]{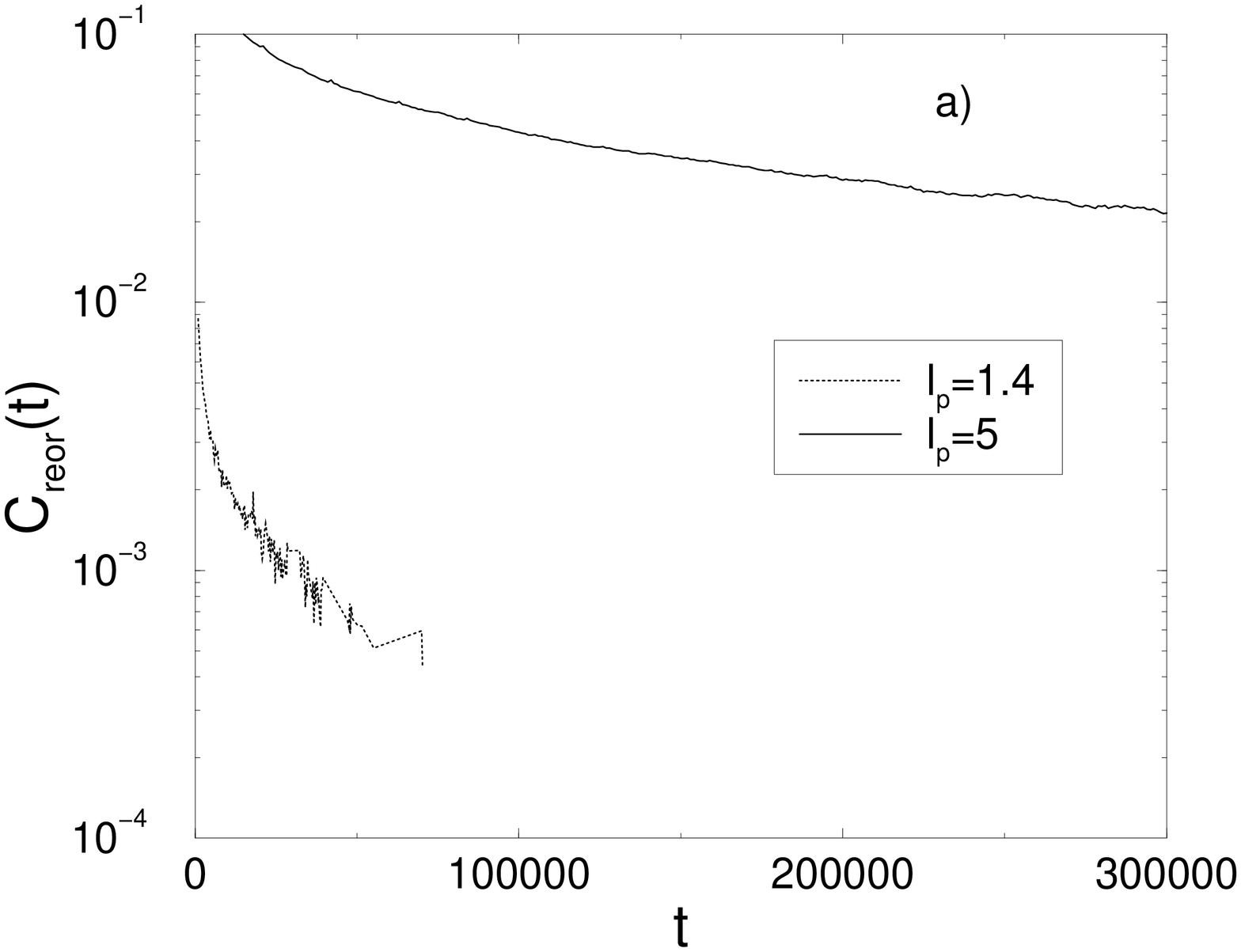}
  \includegraphics[angle=-90,width=0.5\figwidth]{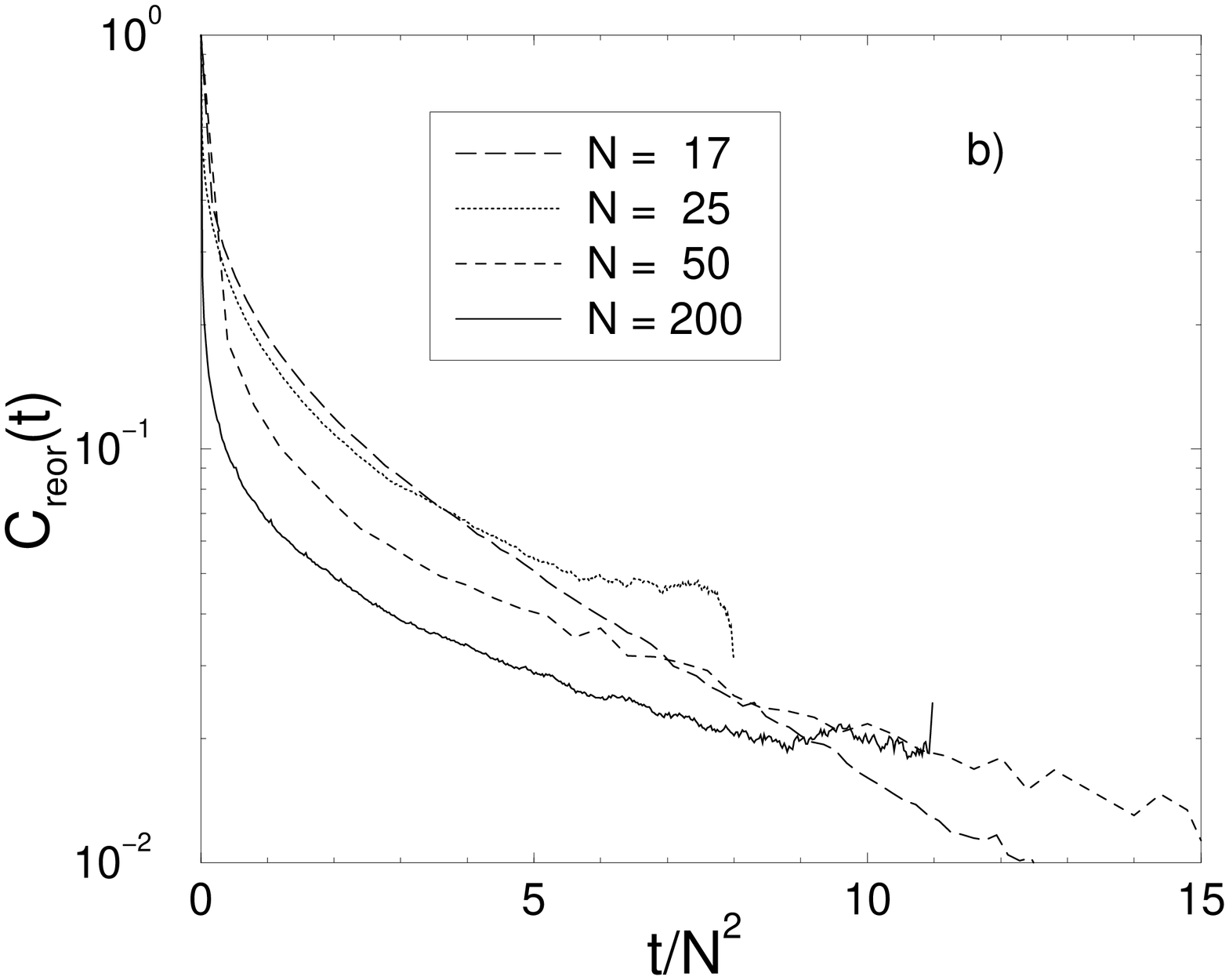}
  \includegraphics[angle=-90,width=0.5\figwidth]{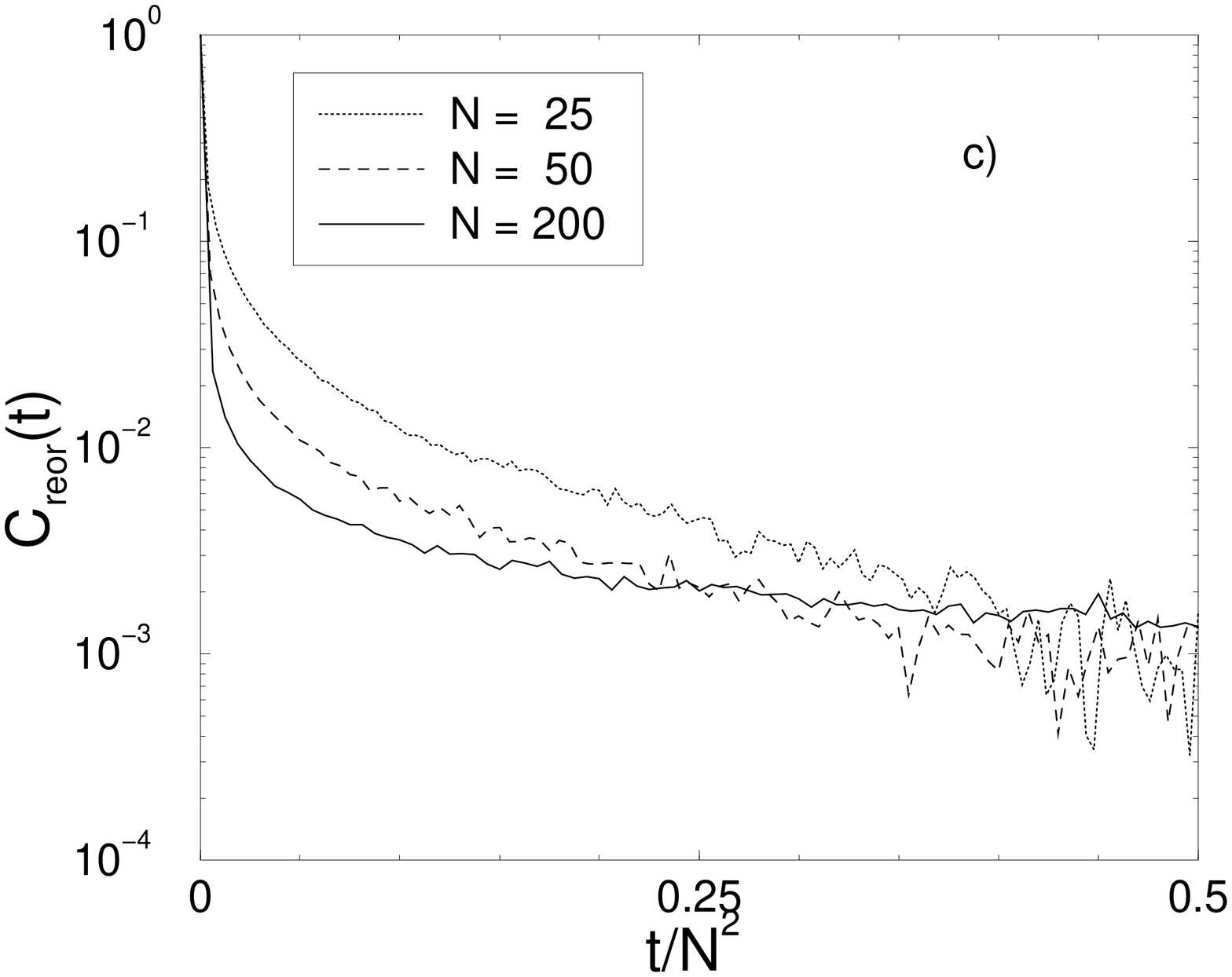}
  \figcaption{Late stage exponential decay of $C_{reor}(t)$ for
    different systems. Time is rescaled by $\frac{1}{N^{2}}$ to show
    differences to Rouse behavior. a)-c) as in Figure~\ref{fig:decay-algebra}}
  \label{fig:exporeor}
\end{figure}
\begin{figure}
  \includegraphics[angle=-90,width=\figwidth]{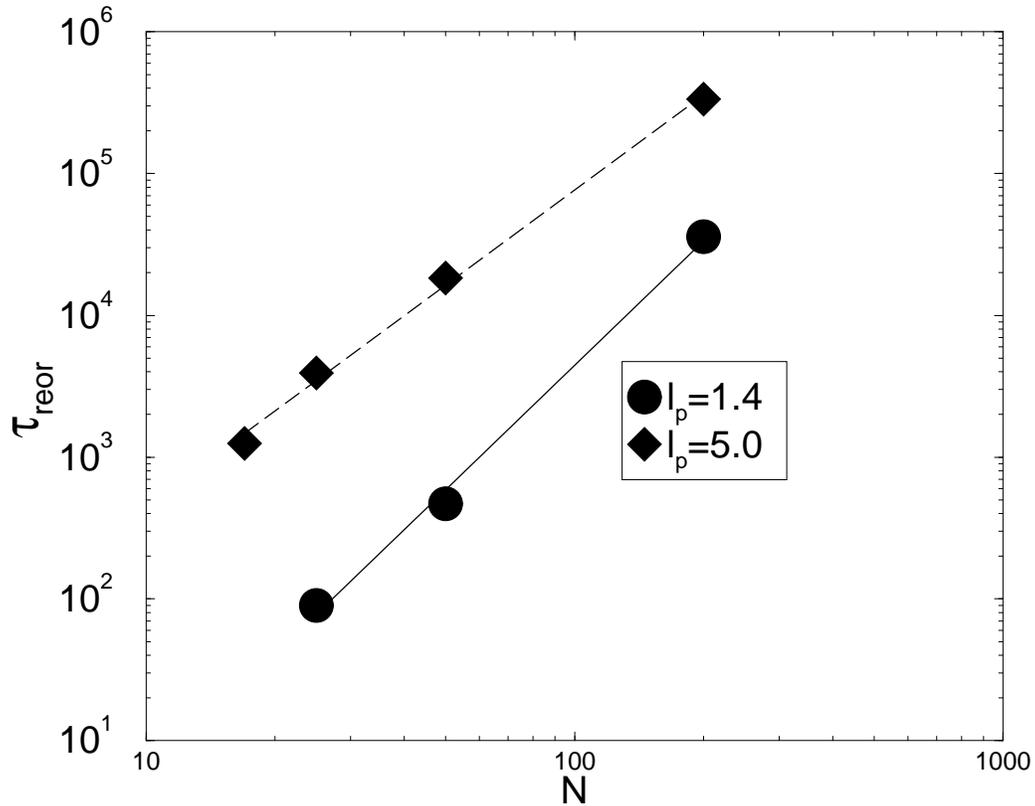}
  \figcaption{Reorientation times $\tau_{reor}$ of local segments as a
    function of chain length and stiffness. These are calculated from
    exponential fits to the long-time part of the reorientation correlation
    function Eq.~\ref{eq:creor}. The dashed line indicates an increase with
    $N^{2.3}$ for $l_p=5$ whereas the solid line indicates $N^{2.9}$ for
    $l_p=1.4$.}
  \label{fig:reortimes}
\end{figure}

The long-time tail of the reorientation correlation function is exponential to
a reasonable approximation (Figure~\ref{fig:exporeor}). The rate of decay
depends~--~as the short-time behavior~--~on both chain stiffness
(Fig.~\ref{fig:decay-algebra}a) and chain length
(Figs.~\ref{fig:decay-algebra}b and c). The reorientation time of the
exponential part $\tau_{reor}$ increases with both chain length $N$ and
stiffness $l_p$ (Fig.~\ref{fig:reortimes}). The dependence on $N$ is stronger
than in the short-time regime, where an approximately linear increase with
chain length was found (Figure~\ref{fig:decay-algebra}d). This is expected
because for larger scale processes the Rouse model should be appropriate and
the overall chain relaxation characterized by the decay of all Rouse modes
should scale with $N^{2}$.  Therefore, the time axis is rescaled in
Figures~\ref{fig:exporeor}b and~\ref{fig:exporeor}c to highlight the remaining
deviation from Rouse behavior. As one approaches the entanglement regime (the
entanglement lengths of the systems are compiled in Table~\ref{tab:entang}),
an even stronger increase with chain length is expected.

The reorientation time $\tau_{reor}$ (Figure~\ref{fig:reortimes})
is much shorter than the time for the reorientation of entire long
chains. Thus, one may suspect that in the case of long entangled chains, it is
no longer important for the reorientation of a tangent vector, whether the
entire chain reorients completely. Instead, the relaxation of a shorter part of
the chain appears to be sufficient.
\begin{table}
  \begin{center}
    \begin{tabular}{|r|r|c|}
      \hline
      $l_p$ & $N_e$ & $\tau_e$\\
      \hline
      1.0 & 32 & 1800 \\
      1.4 & 15 & 1300 \\
      3.0 &  8 & 1300 \\
      5.0 &  6 &  --  \\
      \hline
    \end{tabular}
    \caption{Approximate entanglement lengths $N_e$ (in number of monomers)
      and respective times $\tau_e=\tau_R(N_e)$ for the different persistence
      lengths.These were determined from the mean-squared displacements of
      chains.\cite{kremer90,faller00sa} For $l_p=5$ the entanglement time is
      not defined.}
    \label{tab:entang}
    \end{center}
\end{table}

This is in line with theoretical expectations: In the limit of infinitely long
chains, the local segmental dynamics has to become independent of the chain
length. The relaxation of a large but finite segment of the chain (a few
entanglement lengths long) has to give enough freedom for the local
reorientation to take place. This would only be different if the ends of the
relevant segment were constrained for all times. We were able to corroborate
these considerations by simulating an entangled melt of fully flexible
chains ($N=350,\,l_p=1$) where the initial algebraic decay of $C_{reor}$ is
very fast ($\approx1800t^*$)\cite{duenweg98} and effective in the sense that
it reduces $C_{reor}(\tau_{e})$ to less than 0.01 before the long-time
reorientation sets in. The decay time of the long-time process is about
5000$t^*$ which is the relaxation time of chain-segments of the length of
about 60 monomers. This corresponds to about two entanglement lengths of the
system\cite{kremer90}, so we can deduce that, after completion of short-time
relaxation, only relaxations on a length scale of up to the order of the
entanglement length are relevant.
\subsubsection{Reorientation of medium-size chain segments}
It is also of interest to analyze the reorientation of longer chain segments.
As the orientation vector of a segment of length $d$ we take the unit vector
between two monomers whose indices on the chain differ by $d$
\begin{equation}
  \vec{u}_{d}=\frac{\vec{r}_{i+d}-\vec{r}_{i}}{|\vec{r}_{i+d}-\vec{r}_{i}|}\;.
\end{equation}
Thus $d=1$ represent the vectors connecting neighbors discussed up to now.
\begin{figure}
  \includegraphics[angle=-90,width=\figwidth]{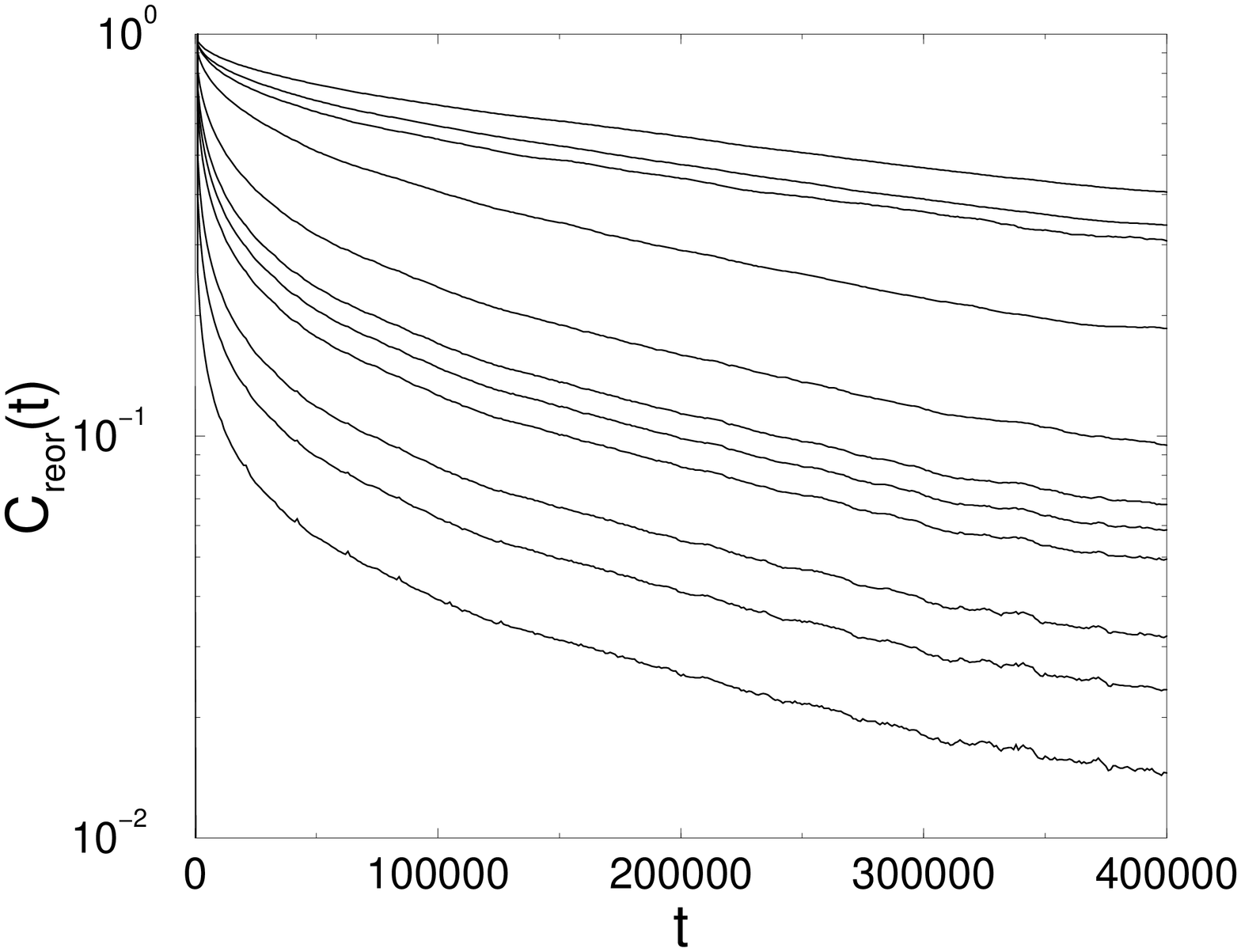}
  \figcaption{Reorientation of chain segments of different length $d$ for the
    system with length $N=200$ and persistence length $l_{p}=5$: $d$=1, 3, 5,
    9, 11, 13, 19, 39, 79, 119 and 199 (end-end vector) from bottom to top.
}
  \label{fig:todfdiffd}
\end{figure}
\begin{table}
  \begin{center}
    \begin{tabular}{|r|r|r|}
      \hline
      $d$ & $\beta$ & $\frac{\tau_{reor}}{1000}$\\
      \hline
       1 & 0.055 & 269 \\
       3 & 0.087 & 272 \\ 
       5 & 0.115 & 276 \\
       9 & 0.172 & 283 \\
      11 & 0.202 & 286 \\ 
      39 & 0.533 & 335 \\
     199 & 0.796 & 558 \\
      \hline
    \end{tabular}
    \caption{Fitted exponential decays to the curves in
      Figure~\ref{fig:todfdiffd} ($N=200,\,l_{p}=5$) in the time domain
      between $t=100000$ and 300000. The amplitude $\beta$ is defined by
      the following relation 
      $C_{reor}(t)\to~\beta e^{-t/\tau_{reor}},\,t\to\infty$,
      i.e. it is the fictitious intersection of the exponential long-time
      curve with the $y$-axis.
      }
    \label{tab:fitdiffd}
  \end{center}
\end{table}
Using simple topological arguments, the dependence of the exponential part of
the reorientation correlation function (Eq.~\ref{eq:creor}) of $u_d$ on $d$
has been predicted.\cite{graf98} The value of the reorientation correlation
function at any given time is proportional to $(3d l_K)/(5 N_el_b)$, where $N$
is the chain length, $N_e$ the entanglement monomer number, and $l_K$ the Kuhn
length. This relation holds for $l_K < Nl_b < N_el_b$.  As a measure of the
amplitude of the exponential part of the reorientation correlation function we
define the ordinate intercept $\beta$ obtained by fitting an exponential to
the long-time tail of the reorientation correlation function and extrapolating
back to $t=0$. If the assumptions of the theory were true, $\beta$ would be
proportional to $d$. In Table~\ref{tab:fitdiffd}, however, $\beta$ is seen to
increase monotonically but sublinearly with $d$. The data of
Table~\ref{tab:fitdiffd} correspond to a non-flexible chain ($l_p=5$). Hence,
it is obvious that the arguments of topological entanglement are not
sufficient to explain the behavior of stiff chains. The reason for this is
most likely that for the $l_p=5$ system $l_K$ and $N_el_b$ are of the same
order of magnitude, as shown elsewhere,\cite{faller00sa} so that the above
condition is not fulfilled.

The long-time exponential parts of the reorientation correlation functions
belonging to different $d$ are almost parallel (Fig.~\ref{fig:todfdiffd}).
They have relaxation times $\tau_{reor}$ of similar magnitude, although
the $\tau_{reor}$ seem to increase slowly but monotonically with $d$ (Table~\ref{tab:fitdiffd}).
\subsection{Comparison to Double-Quantum NMR Experiments}
The direct experimental observable in double-quantum (DQ) experiments like the
ones performed by Graf {\it et al.}\cite{graf98} is
\begin{equation}
  C_{DQ}(t)=\Big\langle P_{2}[\vec{B}\vec{u}(0)]
  P_{2}[\vec{B}\vec{u}(t)]\Big\rangle\;.
  \label{eq:dqformula}
\end{equation}
The vector $\vec{u}$ is a unit vector along an atom-atom distance vector which
is usually not parallel to the chain tangent vector. If
enough of these are available, then the $C_{DQ}$ of the backbone can be
recalculated.~\cite{graf97} $\vec{B}$ is a unit vector parallel to the
external magnetic field in the NMR experiment. We can choose
$\vec{B}=\hat{e}_{z}$ for convenience because amorphous melts are rotationally
invariant.  $C_{DQ}$ is proportional to $C_{reor}$ if the vectors $\vec{u}$
are isotropically distributed with respect to the field
\begin{equation}
  C_{DQ}(t)=\frac{1}{5}C_{reor}(t)\;. \label{eq:dqvsreor}
\end{equation}
Thus, $C_{reor}$ can be used for the comparison to experiments. However,
absolute values cannot be compared, since the experimentally detectable
$C_{reor}(0)$ is reduced from 1 to a value $S$ by very fast motions of
internal degrees of freedom not present in our bead-spring model. For the
C$=$C double bond in polybutadiene, $S$ is found to be 0.24, for
example.\cite{graf98} For comparison, we therefore normalize both curves to
$C_{reor}(0)=1$.

Ball {\it et al.} have derived expressions for the DQ correlation functions
assuming the reptation model and infinitely long chains.\cite{ball97} They
predict an algebraic decay with different exponents in the different dynamic
regimes of the standard tube model. In the time interval between the
entanglement time $\tau_{e}$ and the Rouse time $\tau_{R}$, for which the
inner degrees of freedom of the chain are relaxed, a $t^{-1/4}$ regime of
$C_{reor}$ is expected. Later, in the regime where the chain as a whole
reptates in its tube, a $t^{-1/2}$ behavior should be found. The exponent of
$C_{reor}$ should be the negative of that of the monomer
mean-square-displacement in the same dynamic regime. Algebraic fits of
the short-time part of the decay curves (linear region of the double
logarithmic plot cf. Fig.~\ref{fig:decay-algebra}a-c) yield exponents~$\kappa$
shown in Table~\ref{tab:fitexp}.
\begin{figure}
  \includegraphics[angle=-90,width=\figwidth]{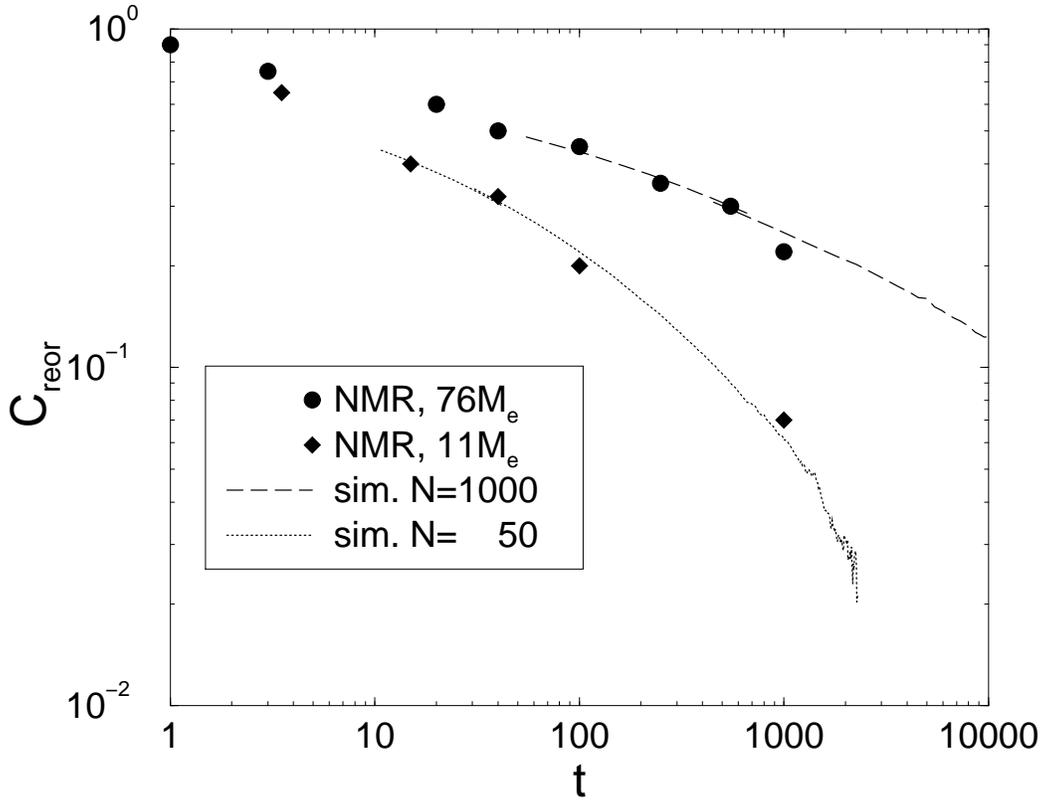}
  \figcaption{Comparison of $C_{reor}$
    from our simulations ($l_p=5$) with the experiments on polybutadiene of
    ref.~\citen{graf98}. Both are scaled to $C(0)=1$. The time axis for the
    simulation data is scaled empirically as an independent  mapping to
    experimental times is not possible. The experimental times are measured in
    entanglement times which are derived from viscosity measurements.}
  \label{fig:dq}
\end{figure}
The systems under study are not very long compared to the infinite chain limit
as they are at most about 30 times the entanglement length (except for
$N=1000,\,l_p=5$). Our exponents~$\kappa$ come therefore closer to the
$t^{-1/2}$ dependence. The system with persistence length $l_p=5$ is the most
strongly entangled.\cite{faller00sa} It is found to reorient slowest with an
exponent between $0.25<\kappa<0.5$. The system with persistence length
$l_{p}=1.4$ shows an algebraic decay faster than $t^{-1/2}$. This is probably
because it is so weakly entangled, that the effects of entanglements are just
starting to play a role. The exponents decrease systematically with
persistence length, which, as discussed earlier, indicates an increasing
degree of entanglement. The dependence on the degree of entanglement is
supported by the very low exponent ($\kappa=0.29$) for a system with $l_p=5$
and chain length $N=1000$.

A $t^{-1/4}$ dependence of the correlation was found experimentally by Graf
{\it et al.}\cite{graf98} for a system with very long and therefore highly
entangled PB chains (76 entanglement molecular weight $M_e$). They also
observed a power-law with $t^{-1/2}$ for a system with a molecular weight of
11$M_e$. Figure~\ref{fig:dq} compares directly simulation (at $l_p=5$) and
experiment; the ratio between the lengths in simulation and experiment is not
exactly the same, it is, however, only important to be slightly or far above
$M_e$. The agreement shows that the simulations reproduce well the exponents
found in experiments. To achieve this agreement, the time axis of the
simulated correlation functions has been rescaled empirically by 0.153 and 0.5
for $N=50\approx8N_e$ and $N=1000\approx160N_e$ respectively. This scaling may
be used to infer a mapping to experimental times.
\begin{table}
  \begin{center}
    \begin{tabular}{|l|l|}
      \hline
      $l_{p}$ & $\kappa$\\
      \hline
      1.4 & $0.68 $\\
      3.0 & $0.50 $\\
      5.0 & $0.40 $\\
      \hline
      5.0$^*$ & $0.29$ \\
      \hline
    \end{tabular}
    \caption{
      Algebraic fits ($t^{-\kappa}$) of the decay of double-quantum
      correlation functions $C_{DQ}$ for $N=200$ (see text and
      Figure~\ref{fig:dq}).
      \hspace{13cm} 
      $^*$ The bottom line for $l_p=5$ has chain length $N=1000$ but the
      system is not equilibrated.
      }
    \label{tab:fitexp}
  \end{center}
\end{table}
\subsection{Interdependence of reorientation and translation of segments}
If the reptation model holds the reorientation process is coupled to the
translation of the polymer in its tube. A useful relation to monitor is,
therefore, the reorientation correlation function of the chain tangent vector
versus the mean-squared displacement of the monomers defining it, irrespective
of the time.  This relation has to be averaged over a finite time window
$2t_{av}$, which is centered at some time $t_m$ and does not necessarily start
at $t=0$.

\begin{eqnarray}
  C_{reor}(\Delta r^{2})&=&\Big\langle 
  P_{2}[\vec{u}(0)\vec{u}(t)]\Big\rangle_{t_m-t_{av},t_m+t_{av}}\;,\\
  \Delta r^2&=&\Big\langle[\vec{r}(t)-\vec{r}(0)]^2
  \Big\rangle_{t_m-t_{av},t_m+t_{av}}\;,
\end{eqnarray}
Both $C_{reor}$ and $\Delta r^2$ depend parametrically on $t$. If,
during~$t$, the tube relaxes (reorients) then $C_{reor}$ is zero. Any
deviation from zero indicates that the orientation is correlated over the time
interval corresponding to the displacement. In our analysis we have ruled out
possible artifacts from the translation of the system as a whole which could
be present in Brownian dynamics.

\begin{figure}
  \includegraphics[angle=-90,width=0.5\figwidth]{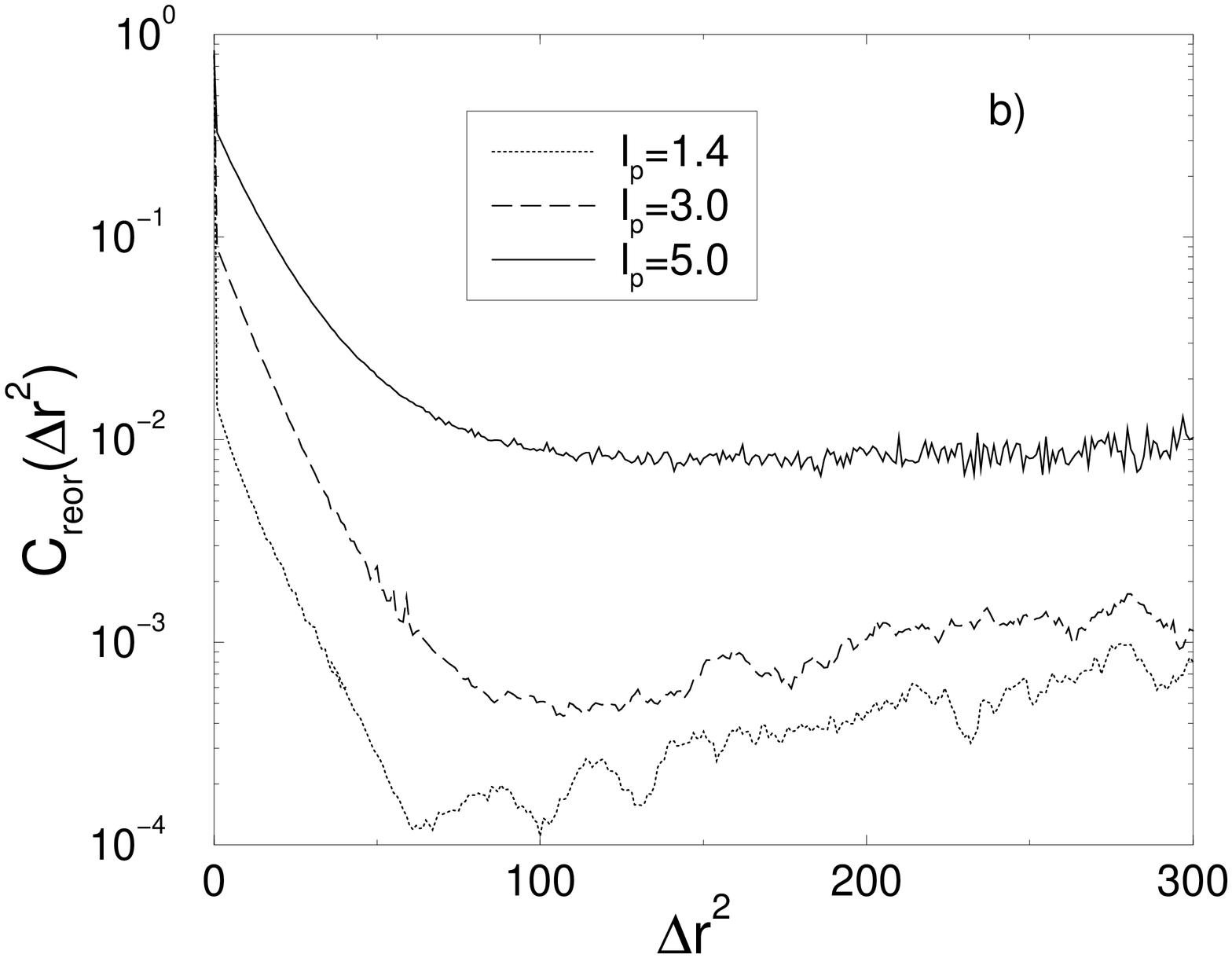}
  \includegraphics[angle=-90,width=0.5\figwidth]{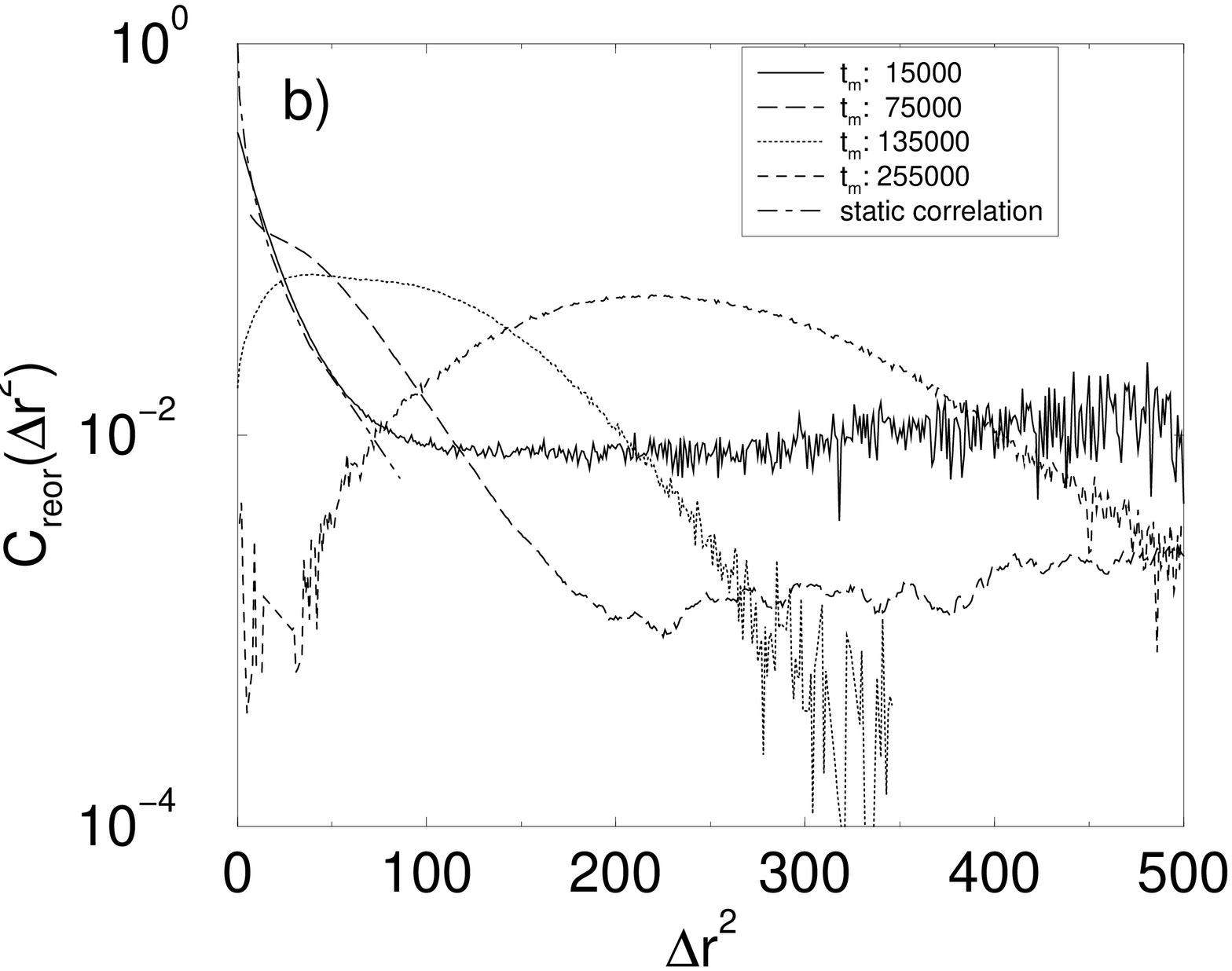}
  \figcaption{Reorientation correlation function depending on the mean-square
    displacement of monomers ($N=200$) 
    a) Different systems at $0\le t\le30000$.  A running average was applied
    after the initial decay in order to show the possible plateau more clearly
    on all the curves in Figure a) and for the curve of $t_m=75000$ in Figure
    b) So the apparent final increase results from low statistics and is not
    meaningful.
    b) $l_{p}=5$ three different time intervals $t_{m}$ and comparison to
    static correlation. For reasons of statistics we always average over
    $2t_{av}=30000$, i.e. $t_m=15000$ means $0\le t_m\le30000$. The static
    case is measured against topological distance, because distance in real
    space is not appropriate as the chain can fold back which leads for
    packing reasons (Section~\ref{sec:structure}) to perpendicular alignments.
    }
  \label{fig:p2ofr}
\end{figure}

In Figure~\ref{fig:p2ofr}a, it is seen that this function at short $t$
($0<t<30000$) does not decay to zero, but shows a plateau, whose value depends
on the stiffness: stiffer chains have a higher residual correlation. The
presence of a plateau is a consequence of the finite length of the chain:
Every finite polymer has a trivial residual static orientation correlation
between distant chain segments in the direction of the end-to-end vector. If
the motion of the chain is predominantly along the fixed tube, this residual
correlation is also visible in the dynamic $C_{reor}(\Delta r^2)$ shown here,
since one chain segment samples the very same fixed tube at different
times. We have seen in the preceding sections that stiffer chains have a
higher reptation component in their dynamics. Hence, it is no surprise that
they exhibit a larger residual correlation.

Figure~\ref{fig:p2ofr}b shows, for the most interesting case $l_p=5$, how the
$C_{reor}(\Delta r^2)$ depends on the position of the time window $t_m$. With
increasing $t_m$, the reorientation correlation at $\Delta r=0$ goes to
zero. At the same time, intensity moves to larger $\Delta r^2$, so that
eventually a maximum at $\Delta r^2 > 0$ develops. This observation is
explained by a scenario that includes, in addition to reptation, a diffusive
or rather subdiffusive translation of entire chain segments through space,
without significant reorientation of these segments: When a chain segment has
reptated along the tube and comes back to its former part of the tube and
hence its former orientation, it finds that this part of the tube itself has
translated in the meantime. On the other hand, when it returns to its former
absolute position ($\Delta r^2=0$) it is now in a completely different part of
the tube and has a different orientation. This picture of short-scale
transverse translation of stiff tube segments has been borne out by
visualizations of individual chains.\cite{faller00sa,faller00a}
The time dependence of the plateau value contains information about the
stability of the initial neighborhood. It measures the ``similarity''
i.e. correlation of the neighborhoods at different points in time as the chain
returns. The neighborhood changes with time on a much longer time scale.

These results support the presence of reptation in our systems, as the chains
come back to their former surrounding which has undergone only small changes
in the meantime. As this memory effect preserves information about
orientations, a tube picture is a suitable concept. However, the chains do not
behave simply as the standard reptation picture would suggest. The reptation
is considerably modified by their stiffness. Stiffer chains reptate in a more
pronounced way, i.e. they follow the primitive path of the tube more closely
as the stiffness suppresses the transversal motions efficiently. This
leads to a higher degree of orientation memory for chains of the same length
(Figure~\ref{fig:p2ofr}a).
\section{Connection to Structure}
\label{sec:structure}
The preceding section has shown that both the stiffness $l_p$ and the length
$N$ have an effect on the dynamics of polymers already on the local level. In
this section, we briefly review earlier results\cite{faller99b} about the
local structure in polymer melts and how it is influenced by the chain
architecture. This is done not only for comparisons within the model
system. NMR experiments on melts have so far only been able to study the local
dynamics. Any information on the structure had to be deduced from the dynamics
using models. In contrast, the simulations of this work can be analyzed
independently for both structure and dynamics. We, therefore, have an example
case for which the assumptions can be checked that are used to analyze NMR
experiments.

Of particular interest has been the question whether or not neighboring polymer
chains are in any way aligned.\cite{kolinski86} We therefore concentrate here
on the static orientation correlation function $OCF$
\begin{equation}
  OCF(r)=\Big\langle P_{2}[\vec{u}_{1}\vec{u}_{2}](r)\Big\rangle\;.
\end{equation}
This function measures the mutual orientation of tangent vectors (defined in
Eq.~\ref{eq:unit}) of segments belonging to two different chains 1 and 2 as a
function of their distance $r$. The second Legendre polynomial is used
again, this time because our polymer chains have no direction, i.e. head and
tail are equivalent. The $OCF$ is~1 for parallel orientation, $-1/2$ for
perpendicular orientation, and~0 for random orientation.

\begin{figure}
  \includegraphics[angle=-90,width=0.5\figwidth]{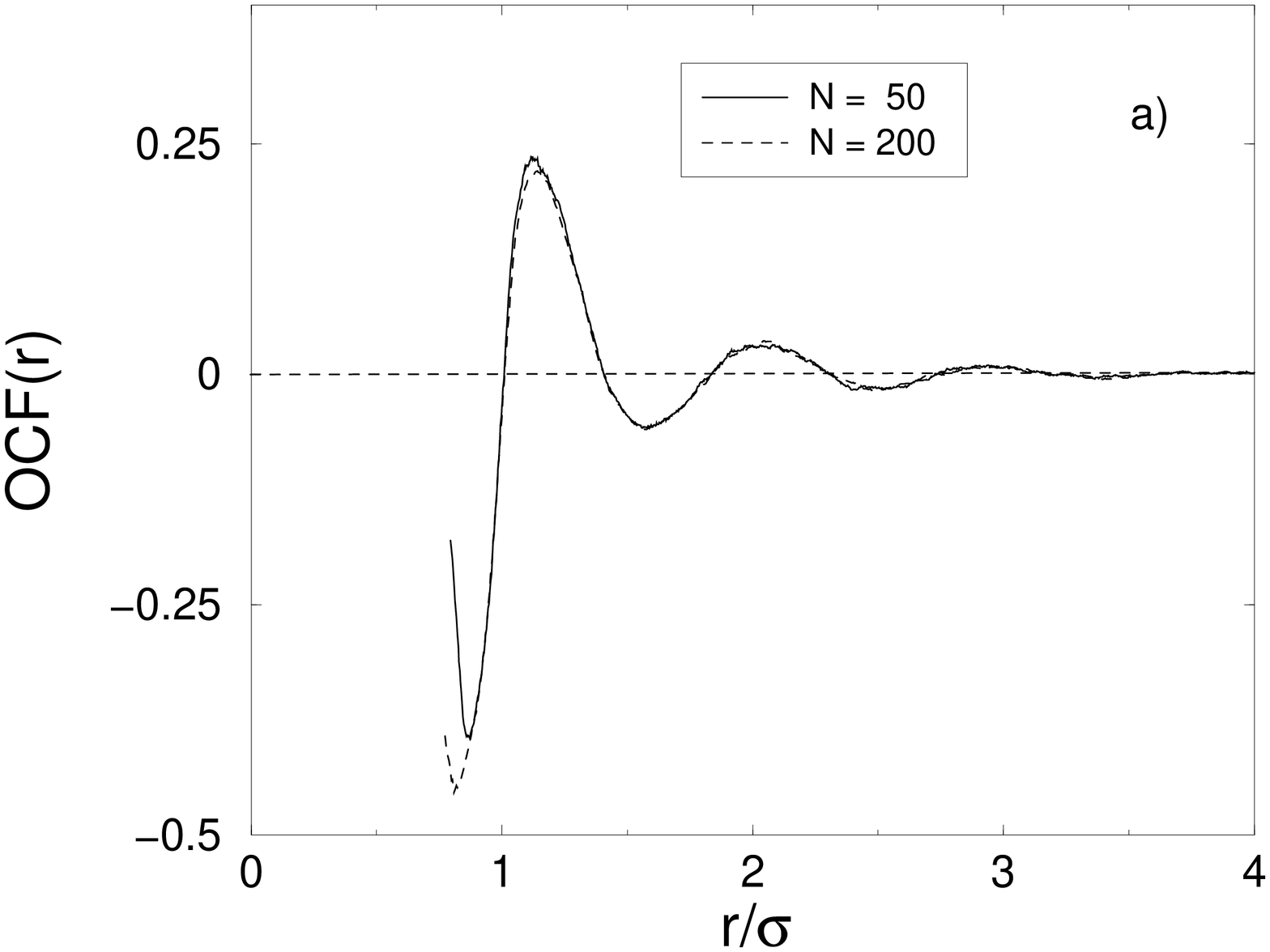}
  \includegraphics[angle=-90,width=0.5\figwidth]{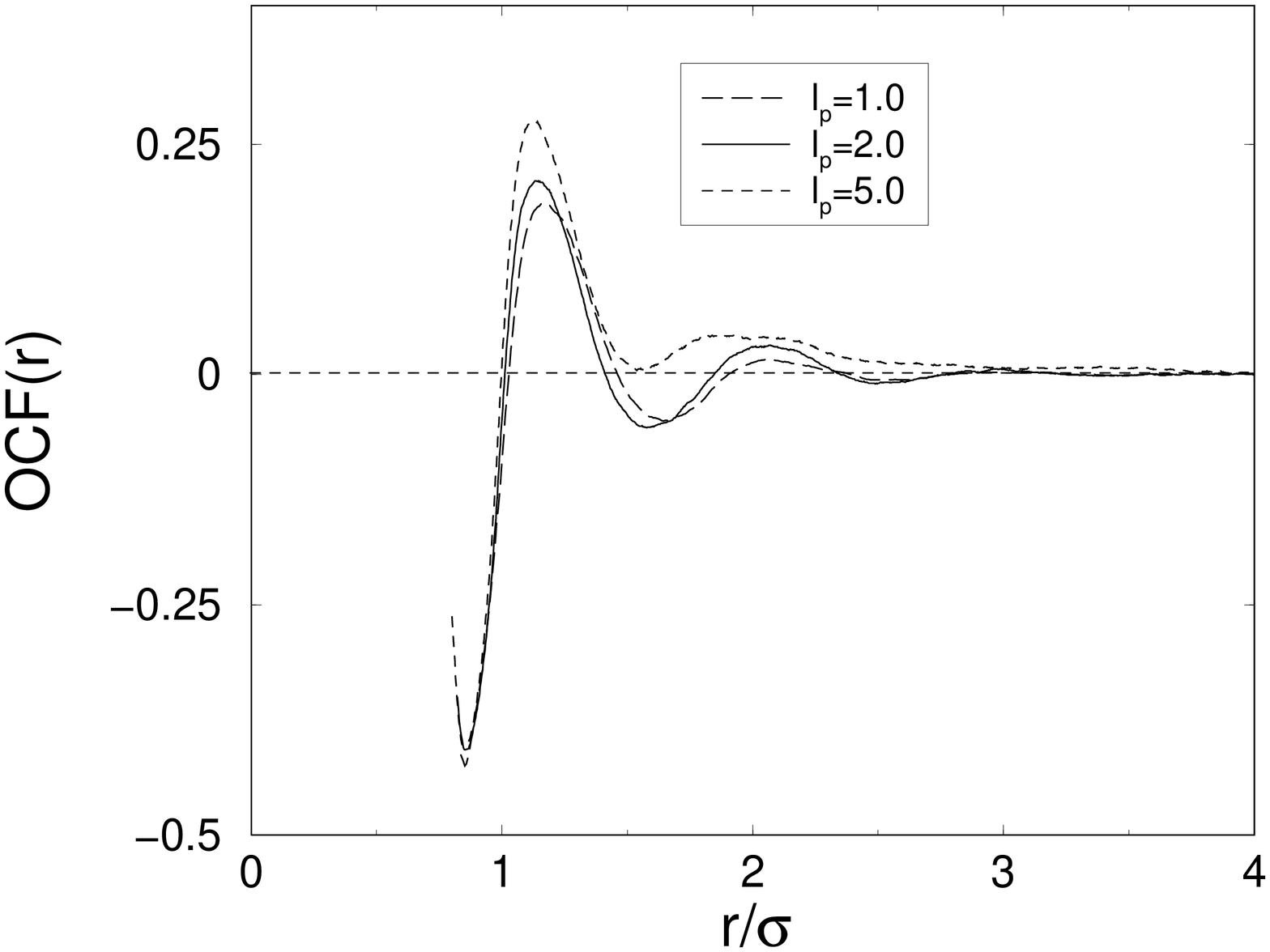}
  \figcaption{Comparison of the static orientation correlation of a) chains of
    different lengths (fully flexible system, $l_p=1$) and b) chains of
    different stiffness ($N=50$) }
  \label{fig:odfdifflen} 
\end{figure}
\begin{figure}
  \includegraphics[angle=-90,width=\figwidth]{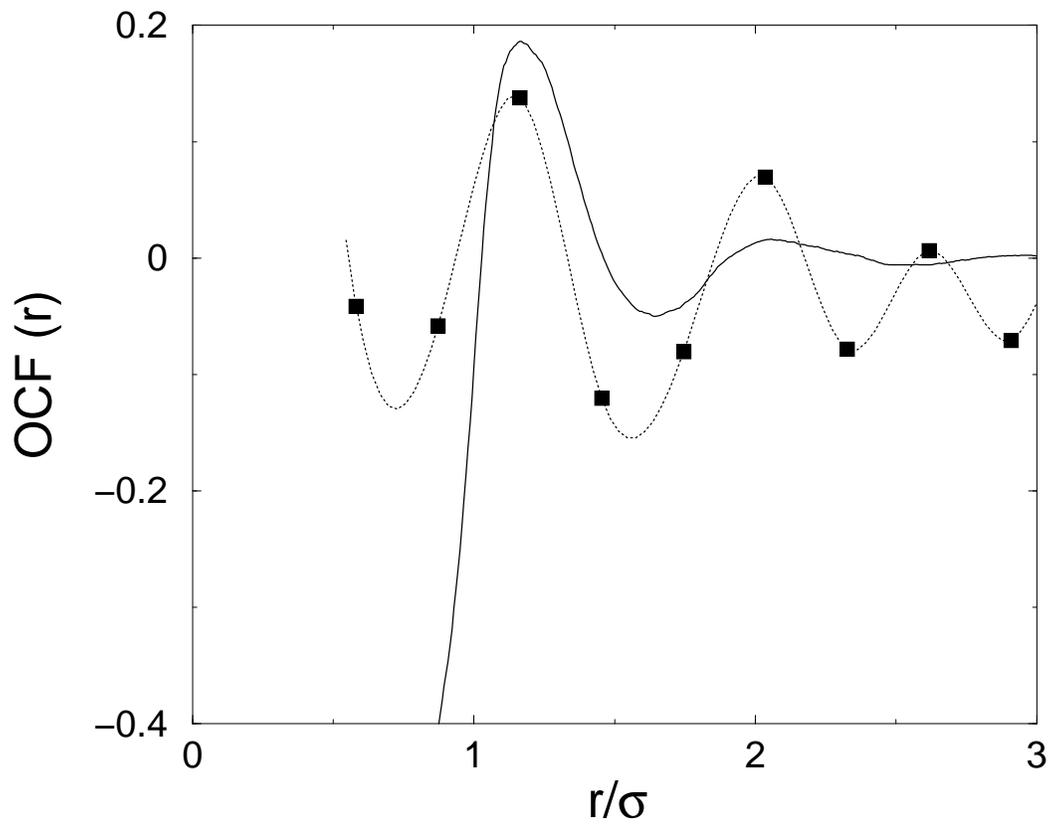}
  \figcaption{Comparison of the static orientation correlation of simple
    bead-spring chains against orientation correlation function of dimers on a
    perturbed fcc lattice, the dotted line is a spline curve as a guide to the
    eye.}
  \label{fig:lattice}
\end{figure}
The detailed discussion of the various $OCF$s is given
elsewhere.\cite{faller99d,faller99b} Here we only note that the $OCF$ is a
strictly local property. The chain length $N$ has no influence whatsoever on
the short-range mutual orientation of two chains, even if one $N$ is below the
entanglement length $N_e\approx32$ and the other above
(Figure~\ref{fig:odfdifflen}a). The influence of the stiffness $l_p$ on the
structure is clearly visible (Figure~\ref{fig:odfdifflen}b), but small. We may
conclude that the two parameters $N$ and $l_p$, which both influence
significantly the local reorientation dynamics, have little ($l_p$) or no
($N$) effect on the local packing of chains. The fact that the chain length is
not important for the local structure means that entanglements cannot be
important either. This is yet another manifestation of the entanglement length
$N_e$ being a purely dynamical quantity.

Local packing is, however, strongly influenced by another local quantity
(which, in contrast, contributes little to the dynamics~\cite{faller00sb}),
namely the excluded volume of the monomers. An example of this is shown in
Figure~\ref{fig:lattice}. Here, the $OCF$ for a flexible chain
($l_p=1,\, N=50$) is overlaid by the $OCF$ for dimers on a randomly
perturbed lattice. The monomers occupy fcc lattice sites whose positions were
randomly displaced by small amounts to emulate finite temperature. The
$OCF$ was then evaluated between all possible pairs of dimers which do not
share a common atom. Although the $OCF$ for the dimers is much more
accentuated than for the amorphous melt one can clearly see the short-range
orientation of dimers shining through in the $OCF$ of the melt.

One may, therefore, conclude that local structure and local dynamics in 
polymer melts are dominated by different properties of the polymer. For
this reason, it may be difficult to infer one from experimental results
on the other.\cite{faller00sb}
\section{Conclusions}
The reorientation of short segments in polymer chains in dense melts is
governed by two subsequent processes. The fast one leads to an algebraic decay
of the reorientation correlation function and the slow one to an exponential
decay.  The correlation times of both depend on the chain length as well as on
the chain stiffness. Increasing chain stiffness leads to a strong slowing
of the reorientation on both time scales.  

A qualitative comparison of our reorientation correlation function showed that
the power laws of the reorientation correlation function measured in
double-quantum NMR experiments of systems not too far above the entanglement
molecular weight could be reproduced.  Therefore, our simplistic model, which
is probably the simplest possible to incorporate stiffness and excluded
volume, successfully describes the qualitative features of the dynamics. With
our results, thus validated against experimental data, the reorientation
correlation functions $C_{reor}(t)$ can be regarded as meaningful. In contrast,
our simple model cannot explain two other experimental observations. The
initial plateau of $C_{reor}(t)$, the so-called dynamical order
parameter\cite{graf98}, of the experiment is not found, and in experiment the
difference between persistence length and entanglement length is
larger. Detailed atomistic models are probably necessary, in order to capture
these features~\cite{faller00a,faller00sb}.

One main result of this investigation is that, for the local reorientation to
relax slowly, the chains have to be both stiff
and long. Local stiffness ensures that the memory of orientation is not
already lost during the fast algebraic process. On the other hand, the chains
have to be above the entanglement length for the slow exponential process to
extend into the experimentally observable regime (milliseconds in
NMR). Although in principle both processes are always present, they have to
span a big enough range in intensity and time, in order to be detectable in
experiments. In agreement with our results, it has been found previously that
the chain length has only little influence on the local reorientation, as long
as it is below the entanglement length. In united atom polyethylene below the
entanglement length local orientation relaxes like a stretched exponential,
whereas the relaxation of the end-to-end vector is well described by the Rouse
model.\cite{harmandaris98}  

Although the dynamic reorientation correlation functions are found to be in
qualitative agreement with experiments and theoretical predictions, our work
shows that this does not necessarily imply an increase of static order due to
topological entanglements of the polymer chains. The static local order
increases with chain stiffness but does not at all depend on chain
length. The entanglement length has emerged as a central quantity to
explain the dynamics of flexible and especially stiff chains. It is not easily
determined uniquely as there are several different definitions which lead to
different values. This difficulty becomes even worse for chains with intrinsic
stiffness. Nonetheless, we can say that the entanglement length in any
definition decreases dramatically as the persistence length is
increased. Evidence for this is found in the chain length dependence of the
center of mass diffusion coefficient and in the segment size governing the
long time relaxation of local order. In the system with $l_{p}=5$, it is
questionable whether {\it any} length scale can be described by Rouse
dynamics. The very local scales up to the persistence are dominated by bending
modes and the very big length scales are dominated by entanglements. As
$N_{e}l_b$ approaches $l_{p}$, the Rouse regime disappears between these two
extremes.

A renormalization of the local scale properties onto an effective monomer or
Kuhn segment is possible for static structural aspects as there is only one
relevant length scale, namely the persistence length. However, this
renormalization fails for dynamical properties of stiff polymers because two
length scales (and the associated time scales) interdepend.\cite{faller00sa}
In dynamics one encounters, in addition to the persistence length, the
entanglement length which describes the topological constraints imposed by
non-crossability of the chains. When the two scales come into the same order
of magnitude, new behaviors emerge which can not be deduced by renormalizing
to Rouse dynamics or other simple models. The fact that the dynamics is not
described appropriately by analytical theories implies also that the
connection of translational and rotational dynamics is not a priori
known. This has to be kept in mind by interpreting NMR experiments which
predominantly measure reorientation.

The concept of reptation is supported by the existence of a
time-dependent plateau value of reorientation as a function of the length of
the diffused path in our simulations. Reptation is more pronounced if the
chains are stiffer because of both the intrinsic stiffness and the stronger
entanglement. They lead to the chain being more closely confined to the
primitive path of the tube.

An analysis of the dependence of $N_{e}$ on $l_{p}$ and further consequences
for chain and monomer motions will be discussed in more detail
elsewhere.\cite{faller00sa} 
\section*{Acknowledgments}
We want to thank M. P{\"u}tz for the long flexible chain data and R. Graf,
K. Kremer and H. W. Spiess for fruitful discussions. Financial support
by the German Ministry of Research (BMBF) through transfer project
``Innovative Methoden der Polymercharakterisierung f{\"u}r die Praxis'' is
gratefully acknowledged.
\appendix
\section{List of Symbols}

\begin{tabular}{ll}
  $\alpha$      & strength of FENE potential\\
  $\vec{B}$     & magnetic field axis\\
  $b$           & strength of bending potential\\
  $\beta$       & amplitude of reorientation correlation function\\
  $C_{DQ}$      & double quantum correlation function\\
  $C_{reor}$    & reorientation correlation function\\
  $D$           & center-of-mass diffusion constant\\
  $d$           & chain segment length\\
  $\epsilon$    & nonbonded interaction strength\\
  $k_B$         & Boltzmann's constant\\
  $\kappa$      & decay exponent for correlation function\\
  $l$           & distance along chain contour\\
  $l_b$         & bond length\\
  $l_p$         & persistence length\\
  $l_K$         & Kuhn segment length\\
  $m$           & monomer mass\\
  $M_e$         & entanglement molecular weight\\
  $N$           & chain length in monomers\\
  $N_C$         & number of chains\\
\end{tabular}

\begin{tabular}{ll}
  $N_e$         & entanglement monomer number\\
  $OCF$         & orientation correlation function \\
  $P_2$         & second Legendre polynomial\\
  $R$           & characteristic length of FENE potential\\
  $R_{end-end}$ & end-to-end distance\\
  $R_{gyr}$     & gyration radius\\
  $r$           & distance\\
  $\vec{r}$     & position vector\\
  $\rho$        & density\\
  $S$           & dynamical order parameter\\
  $\sigma$      & monomer diameter \\
  $T$           & temperature\\
  $T_1$         & spin-lattice relaxation time\\
  $t$           & time\\
  $t^*$         & time unit\\
  $t_av$        & averaging time\\
  $t_m$         & point of time for correlation of orientation 
  and translation\\
  $t_{sim}$     & simulation time\\

  $\tau_e$      & entanglement time\\
  $\tau_R$      & Rouse time\\
  $\tau_{reor}$ & reorientation correlation time\\
  $\vec{u}$     & tangent unit vector\\
  $\vec{u}_d$   & tangent unit vector for segment length $d$\\
  $V$           & interaction potential\\ 
\end{tabular}
\bibliography{standard}
\bibliographystyle{macromolecules}

\clearpage
\parindent0mm
\pagestyle{empty}

\end{document}